\documentclass[a4paper,11pt]{article}
\usepackage{pos}
\usepackage{enumitem}
\title{Gammapy: present status and future roadmap}
 \ShortTitle{Gammapy, a Python package for $\gamma$-ray astronomy}

\author*[a,1]{B. Khélifi}
\author[a]{R. Terrier}
\author[b]{A. Donath}
\author[c]{A. Sinha}
\author[d]{Q. Remy}
\author[e]{F. Pintore}

\affiliation[a]{Université Paris Cité, 
CNRS, Astroparticule et Cosmologie, F-75013 Paris, France}

\affiliation[b]{Center for Astrophysics | Harvard \& Smithsonian,
Cambridge, U.S.A.}

\affiliation[c]{IPARCOS Institute and EMFTEL Department,
Universidad Complutense de Madrid, E-28040 Madrid, Spain}

\affiliation[d]{Max-Planck-Institut für Kernphysik, 
P.O. Box 103980, D 69029 Heidelberg, Germany}

\affiliation[e]{INAF IASF Milano, 20133 Milano, Italy}

\note{~For the Gammapy team}

\emailAdd{khelifi@in2p3.fr}

\abstract{Since its start in 2014, the lightweight open source Python library Gammapy has come a long way to become a popular data analysis package for high-energy astrophysics. Selected as the official CTAO Science Analysis tool, it is also an approved analysis software within the H.E.S.S. and MAGIC collaborations. The first long-term version, Gammapy v1.0 was released on late 2022. It is compliant with several well-established data conventions in high-energy astrophysics, and provides serialised data products that are interoperable with other software. Event lists and instrument response functions curated within the same format from various instruments can be reduced to data binned in energy, time or spatial coordinates. Thereafter, the flux and morphology of one or more gamma-ray sources can be estimated using Poisson maximum likelihood fitting and assuming a variety of spectral, temporal and spatial models. Flux points, likelihood profiles and light curves extractions are supported. Complex user defined likelihoods and models can also be implemented. In this contribution, we will highlight the main features of Gammapy v1.0, including data reduction and analysis examples from different space and ground-based instruments, applications of various background rejection techniques, and a simultaneous fitting across multiple instruments with astrophysical models. We will also present our plans for the future, showcasing new features such as the support of different event types, unbinned likelihood analysis, spectral unfolding and transient source detections. In addition to an improved API with distributed computing for scalable analysis,  enhanced support for all-sky instruments like Fermi-LAT and HAWC is foreseen.}

\ConferenceLogo{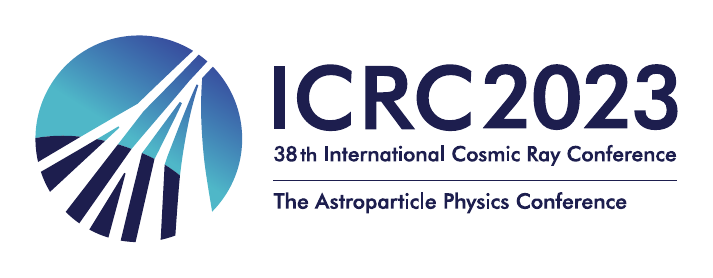}

\FullConference{%
38th International Cosmic Ray Conference (ICRC2023)\\
  26 July - 3 August, 2023\\
  Nagoya, Japan}

\begin{document}
\maketitle

\section{Gammapy, a Python package for $\gamma$-ray astronomy}
\label{sec:intro}

Gammapy\footnote{~Gammapy web site: \href{https://gammapy.org/}{https://gammapy.org/}} is an open Python library~\citep{gpref} aiming to derive astrophysical products and catalogs from very-high energy (VHE) gamma-ray high-level data, such as those from H.E.S.S., MAGIC, VERITAS, HAWC and also Fermi-LAT. These high-level data are produced by \emph{Imaging Atmospheric Cherenkov Telescopes} (IACTs) or \emph{Water Cherenkov Detectors} (WCD) as outputs of their low-level analysis pipelines, consisting of the steps of calibration, reconstruction and background reduction. These high-level data contain event list, Instrument Response Functions (IRFs) and their associated metadata, or also data cubes like an exposure map. Gammapy reads data curated with the {\tt gamma-astro-data-format} (GADF) format~\citep{gadf} and serialised into FITS~\citep{fits}.

\begin{figure}[h!]
\centering
\includegraphics[width=13cm]{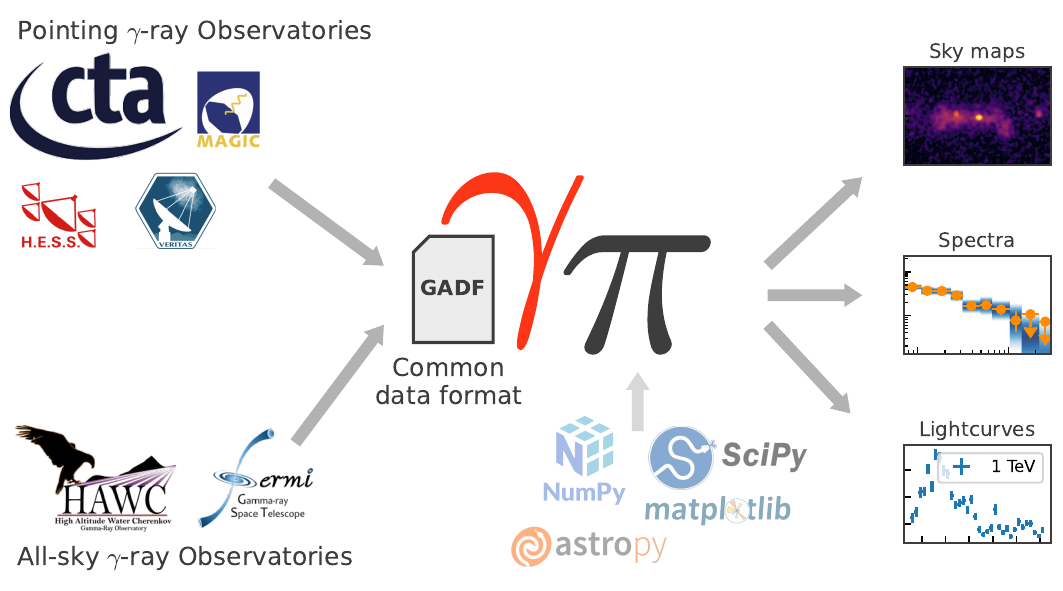}
\caption{Core idea and relation of Gammapy to different $\gamma$-ray instruments and the \emph{gamma astro data format} (GADF).}
\label{fig:gp}
\end{figure}

This package is built within the Python ecosystem and has as main dependencies {\tt numpy}~\citep{numpy} for the n-dimensional data structures, {\tt astropy}~\citep{astropy} (to which Gammapy is affiliated) for the astronomical functionalities, {\tt scipy}~\citep{scipy} for numerical algorithms, {\tt iminuit}~\citep{iminuit} for numerical minimisation and {\tt matplotlib}~\citep{matplotlib} for visualisation. Figure~\ref{fig:gp} summarises the core ideas of Gammapy.

Gammapy is a mature open package built by the VHE community. It has a well defined organisation with Lead Developers, sub-package maintainers and a Coordination Committee, contributors\footnote{~The contributors list can be found \href{https://github.com/gammapy/gammapy/graphs/contributors}{here}.} from different VHE experiments and other domains that are improving the package, the core code, its documentation or its test environment. In parallel, its scientific use is in constant progression\footnote{~List of references using Gammapy: \href{https://ui.adsabs.harvard.edu/search/q=(\%20(citations(doi\%3A\%2210.1051\%2F0004-6361\%2F201834938\%22)\%20OR\%20citations(bibcode\%3A2017ICRC...35..766D))\%20AND\%20year\%3A2014-2023)&sort=date\%20desc\%2C\%20bibcode\%20desc&p_=0}{link}}. In 2021, the CTA Observatory\footnote{~CTAO web site: \href{https://www.cta-observatory.org/}{https://www.cta-observatory.org/}} (CTAO) has selected Gammapy as the core library of its Science Analysis Tool (SAT)~\citep{ctaopr}. It was granted the jury prize in 2022 of the open science awards ceremony of the French Ministry of Higher Education, Research and Innovation~\citep{prize}.

\section{The Long-Term Stable version and the current v1.1 version}
\label{sec:lts}

After the establishment of a roadmap in 2018\footnote{~Gammapy 1.0 roadmap: \href{https://docs.gammapy.org/dev/development/pigs/pig-005.html}{PIG 5}}, the Gammapy team has released in November 2022 its {\bf first Long-Term Stable (LTS) version, the v1.0} (and its minor bug release v1.0.1 in March 2023). An accompagning paper has been very recently accepted by A\&A and is about to be published\footnote{~The paper draft can be found in this \href{https://github.com/gammapy/gammapy-v1.0-paper}{repository}.}. This LTS version will be supported with as many minor bug releases as necessary (following the convention {\tt v1.0.x}), until the release of the next LTS version (see the section~\ref{sec:2.0}). This release is light-weight with about 50,000 lines of code, of which 34\% being the Python code, 22\% being the code test functions, 26\% on the code documentation and 9\% on the package documentation.

\begin{figure}[h!]
\centering
\includegraphics[width=13cm]{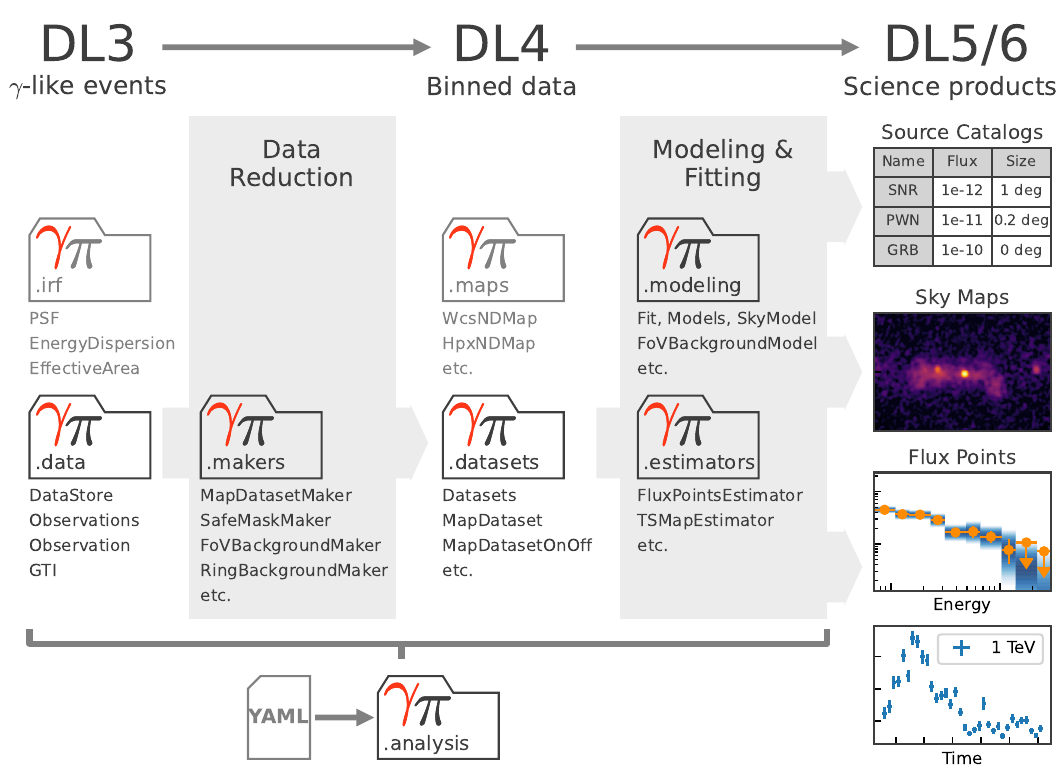}
\caption{Data flow of the Gammapy package: for each data level following the GADF definitions, the container classes are mentionned; the flows between the levels are indicated in grey with their associated classes.}
\label{fig:df}
\vspace{-0.3cm}
\end{figure}

The design of the data flow has been settled, as illustrated in Fig.~\ref{fig:df}. The associated work flow consists of two steps:
\begin{itemize}[noitemsep,nolistsep]
\item the Data Reduction: from user's choices of data selection (or filtering), of geometry (spatial, spectral or temporal) and of analysis type, the IRFs and the event list are selected, projected and reduced into data cubes and stored into a data container class inheriting from the class \emph{Dataset}, that is the central element of Gammapy,
\item the Modeling and Fitting: from user's choices of analysis and its associated parameters, source model(s) is(are) associated to Dataset(s) and their parameters are estimated using Poisson maximum likelihood fitting on the reduced data. 
\end{itemize}

Compared to its previous release (v0.20.1), most of the changes\footnote{~See the \href{https://docs.gammapy.org/1.0/release-notes/v1.0.html}{v1.0 release notes}.} are in the package infrastructure (e.g tutorial gallery relying on {\tt Sphinx gallery}, better compliance with astropy), which permits a simplification of the maintenance. One can highlight that the support for HAWC data analysis has been highly improved, as illustrated in~\citep{hawcgammapy} showing the use of Gammapy by the HAWC collaboration. \\

The package allows to produce different types of VHE astrophysical products using algorithms that have been developed by the TeV community, the GeV community or the X-ray community: 
\begin{description}[noitemsep,nolistsep]
\item[1D spectral analyses] From any spatial region, a spectral model can be adjusted using IRFs made with energy-dependent directional cuts or without and by estimating the background level with the Reflected Background method~\citep{vhebkg}; point-source sensitivity can be computed using ON and OFF estimations; flux points with their likelihood profiles can be derived from any spatial region; 1D spectrum simulation can be performed. 

\item[2D image algorithms] Gammapy can produce maps of significance (using \emph{wstat} or \emph{cash} statistics\footnote{~See the \href{https://docs.gammapy.org/1.0.1/user-guide/stats/fit_statistics.html}{Gammapy statistics page}}) and make source detection; excess maps can be derived by estimating the background with the Ring Background method~\citep{vhebkg}, allowing then 2D map fitting.

\item[Time domain algorithms] Tools permit to extract light curves (flux as a function of time) with different binning strategies, to realise pulsar analysis or to simulate time-variable sources.

\item[Data cube analyses] Gammapy permits to fit spatial, spectral and potentially temporal models on data cubes using background models delivered with the IRFs; flux profile estimations are offered, such as data cube simulations; multi-instrument joint 3D and 1D analysis has been improved.

\item[Global features] Gammapy is delivered with a repertoire of source models\footnote{~\href{https://docs.gammapy.org/1.0.1/user-guide/model-gallery/index.html}{https://docs.gammapy.org/1.0.1/user-guide/model-gallery/index.html}}, that describe spatial, spectral and temporal behaviours: analytical or template models are available; for each fitted parameters, one can estimate their likelihood profiles and also the contour errors between correlated parameters; home-made models or likelihood functions can be used in any analysis; an interface to dark matter spatial and spectral models and to the {\tt naima}~library~\citep{naima} is provided, and also to the open TeV source catalog {\tt gamma-cat}\footnote{~Github repository: \href{https://github.com/gammapy/gamma-cat}{https://github.com/gammapy/gamma-cat}} (that can be visualised also via the web site \href{http://gamma-sky.net/#/map}{http://gamma-sky.net/}).  \\
\end{description}

\begin{figure}[h!]
\centering
\includegraphics[width=8cm]{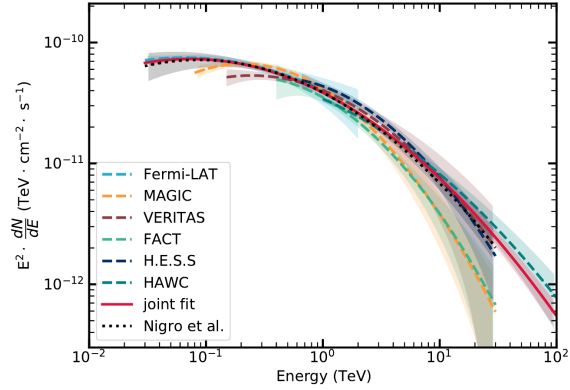}
\vspace{-0.2cm}
\caption{Fig.~13 of \citep{hawcgammapy}: Crab nebula spectral energy dispersion derived with Gammapy.}
\label{fig:jointcrab}
\end{figure}

Since this LTS, a new feature version of Gammapy has been released in June 2023, the {\bf v1.1}. This version has improved the user interface for some classes and functions (the Application Programming Interface, API). So a deprecation warning system has been set up to inform the users. Among the changes\footnote{~See the \href{https://docs.gammapy.org/1.1/release-notes/v1.1.html}{v1.1 release notes}}, a support for energy-dependent temporal models using templates is introduced for the event sampling (ie simulations), and also for multiprocessing. HAWC tutorials have been added, improving our interoperability capabilities. One of the best illustrations (see Fig.~\ref{fig:jointcrab}) is shown by the HAWC collaboration in ~\citep{hawcgammapy} with the joint fit of Crab nebula VHE data from Fermi-LAT, MAGIC, VERITAS, FACT, H.E.S.S. and HAWC to derive a spectral energy distribution with a proper statistical treatment of the data.

\section{The roadmap for the second LTS release (v2.0)}
\label{sec:2.0}

According to the development plan of Gammapy, the LTS v2.0 is targeted for the Fall 2024. A roadmap has been written\footnote{~See the Proposal of Improvement of Gammapy: \href{https://github.com/gammapy/gammapy/blob/b95c80411358df9dec87a0e8628f8327771a4e96/docs/development/pigs/pig-026.rst}{PIG~26}} and details the improvement plan on use cases, features, documentation and insfrastructure:
\setdescription{font=\normalfont}
\begin{itemize}[noitemsep,nolistsep]
\item \emph{event type} handling: for IACTs and WCDs, events can be tagged according to some properties (e.g. their angular reconstruction accuracy, the number of hit sub-detectors) and be associated to specific IRFs in order to improve the analysis sensitivity (e.g.~\citep{ctatype}),  
\item unbinned spectral or 3D analysis: when gamma-ray numbers are limited, for example for transient and flaring sources, unbinned likelihood analysis provides an improvement of the analysis,
\item transient source detection: specific time-domain algorithms to detect variable sources and characterise their variability such as Bayesian Excess Variance,
\item spectral unfolding: extraction of the intrinsic source spectrum with minimal hypothesis on its shape,
\item morphology estimation: tools to measure extension profiles and their associated significances, to determine energy-dependent morphology,
\item handling of systematic effects: from systematic errors stored into IRFs or by adding systematic effects on the IRFs, one could quantify these effects on the final products,
\item nuisance parameters and priors for Bayesian analysis: by adding a systematic effect of unknown amplitude (e.g. a bias in the absolute energy scale), one could estimate the impact of this effect on the parameters estimation assuming a prior distribution of the nuisance parameter.\\
\end{itemize}

The Gammapy team aims also to make significant improvements on the configurable interface of the analysis that uses YAML files, on distributed computing for scalable analysis (using e.g. {\tt Ray}~\citep{ray}), on the 
handling of the future \emph{Very-high-energy Open Data Format} (see Section~\ref{sec:sum}) (clarifying our internal data format and by adding I/O functions), on the handling of meta data and provenance information to respect the FAIR principles for Reseach Softwares~\citep{fair4rs}, on the DevOps tools to facilitate the maintenance and the development of the package (documentation, CI, benchmarks).

\section{The preparation of the CTAO contribution}
\label{sec:ikc}

Gammapy has been selected as the core library of the open Science Analysis Tool of CTAO. The SAT will contain the science tools features of Gammapy and also the functionalities needed for some CTAO operations, such as the real time analysis pipeline. For the CTAO construction, a group of institutions that are strongly involved in Gammapy development and organization have committed to contribute all features to Gammapy to fulfill the minimal requirements of the CTAO as In-Kind Contribution. The acceptance by CTAO of a Gammapy release will follow the CTAO management plan that contains a Preliminary Design Review, a Critical Design Review and an Acceptance Review. 

While maintaining Gammapy as an independent open research software aiming for interoperability between instruments, the Gammapy team is now working in close cooperation with the CTAO computing team. Regular technical meetings permit to define in detail the future SAT features in the context of the global computing needs of CTAO and to collaborate for the improvement of the Gammapy API and documentation. And at the management level, two CTAO representatives are now members of the Gammapy Coordination Committee\footnote{~The composition of the SC can be found \href{https://gammapy.org/team.html\#coordination-committee}{here}}.\\

In the meantime, CTAO is preparing the community to use the future data for scientific purposes. An open Science Data Challenge (SDC) is under consideration with objectives to test some CTAO tools (documentation, data dissemination tools, software distribution), to train the CTA consortium and the wider scientific community to use the SAT to analyse data, and to explore the CTAO expected performances. In this perspective, Gammapy as a core library of the CTAO SAT will be used to produce the simulated data of this future SDC and to analyse the SDC data.

In some regards, this SDC is a stimulating technical challenge for the Gammapy developers. Even if most of the needed features are already present for the data analysis, the developer team is polishing the simulation capabilities with a friendly API, in particular in the time domain. We are also improving the general documentation and the API description, as well as the tutorials. The developers team anticipates an increased participation to schools or to Gammapy hands-on sessions\footnote{~List of the past Gammapy \href{https://github.com/gammapy/gammapy-handson}{hands-on sessions or schools}} and an increased number of general presentations\footnote{~List of the Gammapy \href{https://github.com/gammapy/gammapy-presentations}{general presentations}}.

\section{Summary and conclusion}
\label{sec:sum}

Gammapy is an open Python package for VHE $\gamma$-ray astronomy aiming to derive astrophysical VHE products from high-level data produced by VHE facilities. It uses high-level data curated under a VHE data format, the {\tt gamma-astro-data-format} (GADF). The light-weight Gammapy package is officially in use by several TeV facilities and has been chosen by the CTA Observatory to be the core library of its future open Science Analysis Tool. With the common data format GADF, it permits multi-instrument analyses by supporting joint fit of their data. This mature research software can be used for data from ASTRI, CTA, FACT, Fermi-LAT, HAWC, H.E.S.S., MAGIC, VERITAS. The LHAASO, SWGO, KM3NeT facilities are also testing and evaluating Gammapy. With more than 70~references to Gammapy and with its use for academic purposes, this software grows in popularity and is recognised as a major research software by the VHE community.

The Gammapy team released the first Long-Term Stable version, the v1.0, that offers users a maintained and stable package for their scientific analyses. Its stable user interface associated with minor bug releases offers a comfortable framework for physicists. This LTS offers all the standard and historical VHE algorithms, as well as algorithms coming from other astrophysics domains like the data cube analysis. It also allows a proper statistical treatment of joint analyses of high-level data of different types.

In parallel, the project developers team is working to improve the capabilities of the package, e.g. with the support of Bayesian analysis or unbinned analysis. Until the release of the new LTS, several feature releases are expected and will contain new functionalities and scientific algorithms. It is also expected that the high-level data format will evolve as the GADF initiative has ended. Eleven gamma-ray and neutrino experiments have created a new initiative, called \emph{Very-high-energy Open Data Format} (VODF)\footnote{~VODF web site: \href{https://vodf.readthedocs.io/}{https://vodf.readthedocs.io/}} (see \citep{vodf} for a complete description), aiming to create new standards for VHE astrophysics detectors that respect the FAIR principles~\citep{fair} and follow as much as possible the IVOA\footnote{~IVOA web site: \href{https://www.ivoa.net/}{https://www.ivoa.net/}} standards. In this context, the internal data model of Gammapy is being separated from the read data model with the use of an I/O layer dealing with formats and their versions.

To conclude, the Gammapy team is continuing to improve the open package by offering even more functionalities. As an open project, we encourage any reader to participate to the improvement of the package or to the animation of schools, while taking a special care to recognise any effort. The authors acknowledge again all contributors to Gammapy and all users for their valuable feedback.


%
%
%

\end{document}